\documentclass[11pt]{article}
\usepackage{graphicx,amsmath}
\usepackage[margin=1.25in]{geometry}
\usepackage{authblk}
\usepackage{url}
\usepackage[colorlinks = true,
            linkcolor = blue,
            urlcolor  = blue,
            citecolor = blue,
            anchorcolor = blue]{hyperref}
            
\usepackage{caption}
\usepackage{subcaption}

\usepackage{graphbox}
\usepackage{amssymb}
\usepackage{slashed}
\usepackage{cancel}
\usepackage{xspace}
\usepackage[dvipsnames]{xcolor}

\usepackage{booktabs}
\usepackage{floatrow}
\newfloatcommand{capbtabbox}{table}[][\FBwidth]
\usepackage{multirow}
\usepackage{braket}

\usepackage{alphalph}

\usepackage[]{hyphenat}

\newenvironment{Abstract}{\begin{quotation} \begin{center}
                       ABSTRACT
     \end{center}\bigskip  }{\end{quotation}}

\def\beq{\begin{equation}}
\def\eeq#1{\label{#1}\end{equation}}
\def\eeqn{\end{equation}}

\newenvironment{Eqnarray}%
   {\arraycolsep 0.14em\begin{eqnarray}}{\end{eqnarray}}
\def\beqa{\begin{Eqnarray}}
\def\eeqa#1{\label{#1}\end{Eqnarray}}
\def\eeqan{\end{Eqnarray}}

\let\bar=\overbar

\def\lsim{\mathrel{\raise.3ex\hbox{$<$\kern-.75em\lower1ex\hbox{$\sim$}}}}
\def\gsim{\mathrel{\raise.3ex\hbox{$>$\kern-.75em\lower1ex\hbox{$\sim$}}}}

\def\del{\partial}
\def\Dslash{\not{\hbox{\kern-4pt $D$}}}
\def\dslash{\not{\hbox{\kern-2pt $\del$}}}
\def\pslash{\not{\hbox{\kern-2pt $p$}}}

\def\Dlr{\mathrel{\raise1.5ex\hbox{$\leftrightarrow$\kern-1em\lower1.5ex\hbox{$D$}}}}

\def\MSB{{\bar{M \kern -2pt S}}}
\def\msb{{\bar{\scriptsize M \kern -1pt S}}}

\def\drb{{\bar{\scriptsize D \kern -1pt R}}}

\def\ETmiss{\not{\hbox{\kern-4pt $E$}}_T}

\def\ifb{{\rm fb}^{-1}}

\def\GeV{{\rm GeV}}

\newcommand{\PYTHIA}{{\textsc{pythia8}}\xspace}

\newcommand{\HERWIG}{{\textsc{herwig7}}\xspace}

\newcommand{\HT}{\ensuremath{H_{\mathrm{T}}}\xspace}

\newcommand{\met}{\ensuremath{\cancel{E}_{\mathrm{T}}}\xspace}
\newcommand{\pt}{\ensuremath{p_{\mathrm{T}}}\xspace}

\newcommand{\rinv}{\ensuremath{r_{\text{inv}}}\xspace}

\newcommand{\Ncdark}{\ensuremath{N_{c_\mathrm{D}}}\xspace}
\newcommand{\Nfdark}{\ensuremath{N_{f_\mathrm{D}}}\xspace}

\newcommand{\PZprime}{\ensuremath{\mathrm{Z}^{\prime}}\xspace}

\newcommand\snowmass{\begin{center}\rule[-0.2in]{\hsize}{0.01in}\\\rule{\hsize}{0.01in}\\\end{center}}

\author[a]{Jon Butterworth}
\author[b]{Cesare Cazzaniga}
\author[c]{Aran Garcia-Bellido}
\author[d]{Deepak Kar}
\author[e]{Suchita Kulkarni}
\author[f]{Pedro Schwaller}
\author[g]{Sukanya Sinha}
\author[g]{Danielle Wilson-Edwards}
\author[h]{Jose Zurita}

\affil[a]{Department of Pysics \& Astronomy, University College London, London, United Kingdom}
\affil[b]{ETH Zürich, Institute for Particle Physics and Astrophysics, 8093 Zurich, Switzerland}
\affil[c]{Department of Physics and Astronomy, University of Rochester, Rochester NY, USA}
\affil[d]{School of Physics, University of Witwatersrand, Johannesburg, South Africa}  
\affil[e]{Institute of Physics, NAWI Graz, University of Graz, Graz, Austria}  
\affil[f]{PRISMA+ Cluster of Excellence and Mainz Institute for Theoretical Physics, Johannes Gutenberg-Universit\"at Mainz, Mainz, Germany}  
\affil[g]{School of Physics and Astronomy, University of Manchester, Manchester, United Kingdom}
\affil[h]{Instituto de Física Corpuscular, CSIC-Universitat de Val\'encia, Paterna, Spain} 

\affil[*]{ {\bf Corresponding author(s): Sukanya Sinha \footnote{sukanya.sinha@cern.ch}} }

\date{}

\begin{document}

\title{MITP Colours in Darkness workshop summary report}

\maketitle

\vspace{-60pt}
\snowmass
\vspace{-20pt}
 \begin{Abstract}
{This report summarises the talks and discussions that took place over the course of the MITP Youngst@rs Colours in Darkness workshop 2023. All talks can be found at this URL: \url{https://indico.mitp.uni-mainz.de/event/377/}.}
\end{Abstract}



\section{Introduction}

In recent years, there has been an increase in the number of search programmes exploring the possiblity of a ``dark sector" beyond the Standard Model (BSM) using LHC data. To date, dark matter (DM) searches at the Large Hadron Collide (LHC) have usually focused on WIMPs (Weakly Interacting Massive Particles), but since the standard signatures have found no compelling evidence, several recent phenomenology papers have explored the possibility of accessing the dark sector with unique collider topologies. If dark mesons exist, their evolution and hadronization procedure are currently little constrained. They could decay promptly and result in a very Standard Model (SM) QCD-like jet structure (darkjets), even though the original decaying particles are dark sector ones; they could behave as semi-visible jets (SVJs); or they could behave as completely detector-stable hadrons, in which case the final state is just the missing transverse momentum. Furthermore, depending on whether the dark hadrons decay promptly or not, emerging jet (EJ) signatures can also arise.
 
Owing to the associated experimental challenges, these classes of models are still under developed and mildly explored. Recent developments in reconstruction and identification techniques have made it possible to probe such models at the LHC, and the first limits on some of these signatures are public from both ATLAS and CMS. However, there's still a lot of ground left to cover, in terms of shower and hadronisation approach, and benchmarking the models for future iterations of the searches as we step into an era of unprecedented data, with LHC Run-3 well underway.
 
This workshop~\footnote{\href{https://indico.mitp.uni-mainz.de/e/darkshowers}{https://indico.mitp.uni-mainz.de/e/darkshowers}} aimed to foster collaboration between the experimental and theory community dedicated towards developing and understanding the strongly interacting dark sector. The workshop featured talks from the leading experts in the field, which were followed by extensive discussion sessions, to understand the current status of the dark showering module within Monte Carlo generators like \PYTHIA and \HERWIG, as well as identifying potential studies pertaining to signal generation and designing theoretically motivated models that will drive future search strategies for strongly interacting dark sectors. 
\section{Day 1}

The day focused on the theoretical perspective, and on the status of event generation. All theory speakers reminded us of the vast landscape of phenomenological scenarios and their signatures, and in all cases the discussions focused on how to improve the search strategies to capture a broader set of scenarios than currently aimed for.

The landscape of Hidden Valley (HV) scenarios~\cite{Strassler:2006im,Han:2007ae,Juknevich:2009ji} is vast~\footnote{Presentation by Matthew Strassler}. The signatures can be classified in the ``easy'' (parton level MC suffices), ``feasible'' (a dark shower Monte Carlo is needed) and ``guesswork'' (no reliable simulation). For the ``easy'' case, the reach can be broadened by making searches as inclusive as possible, for instance by relaxing event selection criteria. This point was illustrated using the 
$Z \to A' S, S \to A' A'$ (where $A'$ is a dark photon, $S$ a new scalar and $A'$ decays to leptons) search done in~\cite{ATLAS:2023jyp}, proposing to replace the lepton isolation criteria by a displacement cut, as often non-isolated leptons would arise in the decay of dark jets. On the “feasible” signatures, the current QCD knowledge can be exploited to explore this region better. Concretely, \PYTHIA can simulate the simplest case of i) perturbative (QCD-like) theories, ii) with mass degenerate dark quarks and iii) with all dark quark masses lying below the dark confinement scale $\Lambda$. Deviating from any of these three assumptions can render \PYTHIA unreliable and/or not applicable, and theory/simulation progress is required to improve the situation.

In order to simulate confining HV theories and analyse their phenomenology, development of event generators is necessary~\footnote{Presentation by Suchita Kulkarni}. The main developments to the \PYTHIA HV module during and since the Snowmass process~\cite{Albouy:2022cin} were highlighted. These include the possibility to abandon the mass degeneracy hypothesis, the implementation up to three loops of the dark sector coupling constant up and allowing the SM Higgs decay into dark gluons $\texttt{gv}$. These new features must be analyzed carefully, as \PYTHIA raises only few sanity warnings. Possible improvements to the HV module are the expansion to include other Lie Groups beyond the $SU(N)$, and considering that the dark “quarks” must not necessarily be Dirac fermions. The importance of estimating hadronization uncertainties away from SM QCD point was emphasised. 

Among the experimental signatures of confining HV scenarios emerging jets ~\cite{Schwaller:2015gea} are an interesting phenomenon~\footnote{Presentation by Pedro Schwaller}. While only the simplest case has been study, considering more general assumptions, like i) a mixture of different dark pions lifetimes, ii) more than one dark meson, open up new avenues. On i), the reinterpretation of current EJ CMS search for the case of two different dark hadron lifetimes was demonstrated~\cite{Mies:2020mzw}, while on ii) interesting phenomenology can be obtained from the “down” portal and “up-portal” (mostly focused on “top portal”) scenarios depending on the flavour structure of the dark sector-SM interactions\cite{Renner:2018fhh,Carmona:2021seb}, which can lead to exotic top decays, $t \to u X$, where $X$ can be either a (long lived) dark pion or result in a full dark shower~\cite{Carmona:2022jid}. Moreover, scenarios like ii) can also lead to consistent models of dark matter, where the relic density is calculable. Finally it was emphasised that by working with concrete benchmark models, the reach of collider searches can be compared with e.g. cosmological constraints (CMB, BBN) and with the reach of fixed target or flavour probes of dark sectors. 

Although HV scenarios offer theoretically and experimentally interesting landscape, the number of free parameters may introduce model dependence. Thus an alternative framework for simulating confining HV scenarios was discussed where both emerging and semi-visible jets can be accommodated~\footnote{Presentation by Nishita Desai}. A brute-force approach would require choosing too many parameters in \PYTHIA, and hence she investigated the impact of them on the observable energy distributions. The proposal involves a simplified setup using 6 parameters: $R_{\rm max}$ (Jet Radius parameter), $a$ (Jet energy shape), $N_H$ (avg \# of dark hadrons), the fraction of invisible hadrons \rinv, the mass of the dark hadron $m_H$, and the decay table of dark hadrons. This has been provisionally validated for the \PZprime mediated dark showers.  

Finally, the implementation of dark showers in the \HERWIG framework was also discussed~\footnote{Presentation by Dominic Stafford}. While many aspects are still undergoing development (in particular the hadronization modules), dark splitting in shower module has been successfully implemented, and preliminary results show similar production rates than those obtained with \PYTHIA. 

\subsection*{Discussion summary: Day 1}

In summary, the main challenge on the theory frontier is to capture the vast range of possible signatures while taking into account the limits of theoretical knowledge, availability of simulation tools, and the finite number of phenomenological studies and experimental searches that can be done in practice. The progress on simulation tools, simplified phenomenological models and simple but complete benchmark models reported above looks promising and makes us confident that the challenge can be overcome in the future. 
\section{Day 2}

Day 2 of the workshop was mostly focused on critical review of experimental results and possible future improvements, prefaced by an introductory theory talk on SVJ. The initial model was designed to give rise to the desired topology, with a limited set of parameters in \PYTHIA HV module. However it is clear that these parameter choices result in certain amount of model dependence~\cite{Cohen:2020afv}. 
 Semi-visible jets production have been considered so far mainly in the context of Simplified Models ($s$-channel and $t$-channel), as well as portal contact interaction coupling SM quarks with dark quarks, later undergoing showering and hadronization and decay to the SM sector. More recently, an alternative production mechanism of SVJ via glueballs was suggested in \cite{Batz:2023zef}\footnote{Presentation by Tim Cohen}. Furthermore, in the direction of better understanding the jet substructure and having control of the model-dependence, recently the Lund Jet Plane (LJP) approach~\cite{Dreyer:2018nbf} has been applied to dark showers~\cite{Cohen:2023mya}. The main idea of the proposed method relies on the fact that the non-perturbative effects are isolated in the lower region of the LJP (below $\ln k_t < \ln \Lambda_d$). Thus, one can have control on the impact of the non-perturbative effects removing emissions in the LJP below a certain $k_t$ threshold.

CMS presented their early Run 2 emerging jets results~\cite{CMS:2018bvr}\footnote{Presentation by Jannicke Pearkes}, which uses the impact parameters of tracks in jets to separate signal regions. The b-jet contamination in the background acceptance was carefully studied. The full Run 2 result is coming soon. Run 3 analysis is expected to use dedicated triggers and a ML approach to tag the jets, potentially combining with SVJ. CMS then reviewed the full Run 2 $s$-channel SVJ search~\cite{CMS:2021dzg}\footnote{Presentation by Aran Garcia-Bellido}, motivating the specific HV parameter choices made. Cut-based and BDT jet-tagger strategies were used, and new filters were developed to discard problematic regions of the detector artificially enhancing the data sample with events characterized by \met aligned with the second leading jet, data of interest for the search. The next searches will focus on $t$-channel SVJ production, boosted topology, and uncovered phase spaces, such as lower mediator masses and SVJs with leptons. 

Then ATLAS presented their $t$-channel SVJ \cite{ATLAS:2023swa}\footnote{Presentation by Deepak Kar} and $s$-channel darkjets~\cite{ATLAS-CONF-2023-047} results\footnote{Presentation by Dilia Maria Portillo Quintero}. The former used a \met trigger as opposed to the un-prescaled jet pt and \HT used in the CMS search. This could have allowed to probe intermediate values of \met but the analysis did not because of the difficulty in fake multijet modelling, which will be an useful aspect to study going forward. The darkjets search mainly used number of tracks in ungroomed jets to identify signal but the usual issue is de-correlating it from large multijet background. 
While experimentally track multiplicity is an useful variable, more robust observables, such as subjets inside jets~\cite{Strassler:2008fv}, their correlation and masses can be useful as probes of meta-stable hidden hadrons.
The coupling was chosen such that the models have not already been excluded by existing dijet searches. 

The ECS talks highlighted the HEPData~\cite{Maguire:2017ypu} preparation for this result\footnote{Presentation by Danielle Wilson-Edwards}, which is an important input to make this result usable by larger community, as well as potential improvements in these searches by using partial event building\footnote{Presentation by Angelica Aira Araw Ayalin}, allowing us to expand the search coverage. An attempt to simulate dark showers for near-conformal confining Hidden Valleys, using the \PYTHIA~\cite{Sjostrand:2014zea} HV module, was also presented\footnote{Presentation by Joshua Lockyer}.

\subsection*{Discussion summary: Day 2}

The discussion session focused on possible benchmark signal choices.  Model-dependence can arise at different stages when considering the different unconventional jets production: from dark quarks production going to dark showering, dark hadronization and decays. However, fixed the portal and given the assumption of a QCD-like hidden sector, as well as choosing certain number of dark colours, \Ncdark and dark flavours, \Nfdark allows some control on the perturbative steps of the process. The modeling of the non-perturbative regime of these QCD-like hidden sectors (hadronisation and dark hadrons spectrum) remain one of the main unknowns that can strongly impact our predictions for a scenario where the presence of a new confining dark sector leaves its imprint on the substructure of QCD-like jets~\cite{Cohen:2020afv}. This aspect is extremely relevant for searches exploiting this powerful information. 

The choice ATLAS made to go with $\Nfdark=1$ (as opposed to $\Nfdark=2$ in CMS) has certain conceptual difficulties, but it still generates the topology of interest. It was mentioned that choosing $\Nfdark > 1$  resolves the issues pertaining to dark meson production, and $\Ncdark > 2$ can help to avoid problems associated with hadronization and baryon production, however, then it becomes challenging to implement the \rinv fraction consistently. The suggestion was to force all pions to be either invisible or visible, and set decay width by hand, but even then, simulating mass-split theories where cascade decays within the dark sector can lead to more ambiguities. There are alternate approaches like using a Gaussian smearing, discussed in the previous day of the workshop, to avoid complicated HV parameterisation, however further studies are required on that front. It was also concluded that the \PYTHIA 8 HV module is currently insufficient to describe the expected behaviour for near-conformal confining Hidden Valleys, due to the presence of infra-red fixed points (IRFPs). However, the studies show significant progress on designing a framework which can adequately describe the behaviour at high \Nfdark. 

Independent tests are needed to gain more confidence in the 
different classes of theories that can be studied using this approach.
It is a very much an open issue, and a compromise has to be struck between theoretically 
sound models and experimentally searchable models focusing on uncovered phase space.

\section{Day 3}

The day began with discussion of complementarities between the dark sector and DM, and how simple DM limits maybe interpreted in more complex
scenarios\footnote{Presentation by Mark Goodsell}.
There are indications from simulations of galaxy formation that models in which DM is a single species of particle with negligible self-interaction (apart from
annihilation in the very early universe) struggle to describe the density curves for galaxies, producing a ``cusp'' in density rather that the oberved broad
core~\cite{Spergel:1999mh}.
While interactions between baryonic matter and DM may be able to explain this~\cite{Oh_2011}, self-interactions amongst DM also
provide a potential explanation~\cite{Rocha:2012jg}.
In this case, the strength of self interactions required can be obtained from scales similar to $\Lambda_{\textrm{QCD}}$, and
thus strong interactions become potentially relevant. 

For DM production in such scenarios, there are several mechanisms, generally involving either kinetic mixing between a dark state and the photon and/or Z boson,
or the introduction of a scalar mediator coupling to or mixing with the SM Higgs boson. Freeze-out of the appropriate DM relic abundance
can be obtained by exploiting $3 \rightarrow 2$ processes~\cite{Hochberg:2014dra} or by using heavier states such as the dark $\rho$
mesons~\cite{Bernreuther:2019pfb}. When the $\rho$ lifetime is long, this leads to semi-visible jet and displaced vertex signatures at the LHC.

Standard tools such as Micromegas~\cite{Belanger:2004yn} are not (yet) equipped for computing such non-standard relic density channels, and such calculations
are done piecemeal.
In addition, given the complex phenomenology of these kinds of models and the wide range of possible experimental signatures, it is not efficient to design
individual collider search strategies for every scenario, and reinterpretation of experimental results in multiple scenarios becomes critical.
This is feasible, and has been done, for dijet and monojet analyses, but is a big challenge for the more unconventional signals
such as semi-visible jets and long-lived particles. Sometimes this is due to unavailability of detailed experimental information, especially as
complex machine learning tools are employed. Agreeing on how to share such information between experiments and with the theory community is a critical issue.

For the more standard final states at colliders, one approach is to consider particle-level measurements\footnote{Presentation by Jon Butterworth},
using the Contur package~\cite{Buckley:2021neu,Butterworth:2016sqg} which uses the analyses stored in Rivet~\cite{Bierlich:2019rhm} to perform
a signal-injection of events predicted by a specific BSM scenario on a library of hundreds of fiducial cross section measurements. If the signal would
have been seen, the model point can be excluded. If the expected and actual limits begin to diverge, with the expected limit stronger than the actual,
this may be a sign of an anomaly.
 
To be ideally usable in Contur, a measurement needs to be unfolded to a ``particle level'' fiducial cross section, that is, corrected for detector effects such as
resolution and efficiency within a kinematic region of good acceptance, and not extrapolated beyond that region, since such extrapolations (including correcting
for vetoes on reconstructed objects) inevitably introduce model dependence. It is also better, though not essential, that the measurement is defined in terms
of the true final state, not in terms of production processes. Hundreds of LHC measurements already meet enough of these criteria to be usable, and are
present in the Rivet and HEPData.

Several DM models have been tested by Contur, including a strongly interacting Dark Matter model~\cite{Kribs:2018ilo,Butterworth:2021jto,ATLAS:2023vgr}, and a combined study using an interface to GAMBIT~\cite{GAMBIT:2023yih}, but these are all limited to
models where the final state consists of (potentially novel) combinations of SM objects. Extending this to objects such as non-isolated leptons and
photons, emerging jets or long-lived particles is a challenge, but in general if an object definition can be made in terms of final state particles, and
implemented in Rivet, Contur can make use of it.

This discussion led naturally on to a discussion of analysis preservation for consumption both inside and outside collaborations, particularly as
applied to strongly interacting DM models\footnote{Presentation by Louie Corpe}.
This is to some extent a solved problem for SM-like measurements, where many are already available in Rivet.
Once this information is made available, theorists can reinterpret the searches. Similar tools are available for searches,
and there are community guidelines and discussion documents available~\cite{LHCReinterpretationForum:2020xtr,Bailey:2022tdz}.
Object or event selections using complex machine learning methods currently pose a challenge, especially when they
mingle particle- and detector-level concepts.
Detector level simulations for theorists are difficult to implement, due to the unavailability of collaboration-backed simulation configurations,
and the limitations of fast detector simulations and parameterisations. 
Many discussions are underway as to how to meet this challenge, and there is an urgent necessity for feedback on the usefulness of the available
experimental information for reinterpretation.
To ensure the usefulness of data well beyond the lifetime of the experiments, object definitions have to be made clear from the experimental side at some
level which is interpretable in terms of theoretically-accessible objects, ideally final-state particles. These challenges for non-standard objects are not,
however, unique to unfolded measurements, and have for example to be met (and have been met) even for the calibration of objects used in detector-level
searches. The final step to making them reinterpretable is hopefully therefore not insurmountable.

Then followed a series of presentations by early career researchers and discussions of specific analyses. 
A reinterpretation of CMS emerging jet search in the context of exotic Higgs boson decays~\footnote{Presentation by Juliana Carrasco}, following~\cite{Carrasco:2023loy}, demonstrated a closure test using acceptances provided for CMS signal benchmark points, showing that the search was reproducible using information
available on HEPData. Different exotic higgs decay portals were then checked, and bounds were set which are better than the current best bounds (from ATLAS).
Such studies can also show where are limits are less strong, and thus lead to proposals for improvement.


A study of semi-visible jets dominated by b-quarks was presented~\footnote{Presentation by Wandile Nzuza}, focusing on a search strategy that utilizes variable radius jets to better encompass the semi-visible jet behaviour on an event-by-event basis \cite{Kar:2022hxn}. There are some constraints on the final state from current searches that probe b-jets, however, there is sufficient phase space left to be explored. The current CMS semi-visible jet search exploits the SVJ b-enriched content via a BDT-based jet tagger. Constraints from this search should be compared with this studies, together with constraints from the current ATLAS SVJ jet search.

Semi-visible jets can also be produced with non-isolated lepton pairs being present inside the jet, and there were several presentations focusing on the case of non-tau leptons within SVJ~\footnote{Presentation by Cesare Cazzaniga}, as well as having Non-isolated taus in SVJ~\footnote{Presentation by Tobias Fitschen} based on \cite{Cazzaniga:2022hxl,Beauchesne:2022phk}.
No strong bounds from existing searches are present for either scenarios. Potential search strategies can include lepton counting or mass correlation between the hardest lepton pairs.
Topological triggers tend to be more promising for such a final state compared to standard triggers, and alternative triggering strategies such as Trigger-Level Analysis, or Scouting, accompanied by Partial Event Building (PEB) can be explored for both final states. The signal generation for SVJ with leptons can be approached via different methods, as discussed in the last presentation ~\footnote{Presentation by Clarisse Prat.}.

\subsection*{Discussion summary: Day 3}
The different presentations throughout the day triggered some discussion about how to best preserve existing analyses, and make them accessible and/or reproducible for the wider community beyond the respective experiments. The theory community came up with a wishlist of information that would help them in reinterpreting relevant analyses in the context of new models. Full likelihoods seem to be a mandatory requirement, with additional information coming from covariance matrices, explicitly spelling out intermediate steps in the analysis chain and descriptive object definitions. It was mentioned that the efficiency maps provided tend to be model dependent and hence difficult to reinterpret. Existing packages can be utilised better, i.e. providing validated SimpleAnalysis~\cite{ATLAS:2022yru} routines, REANA~\cite{Simko:2018zzz} implementations of the full analysis workflow, or using the detector smearing available within Rivet~\cite{Buckley:2019stt}. The cutflow tables provided by experiments are a good starting point, and attention to detail is required when adding information to HEPData. There was an overarching demand for analyses that use sophisticated ML techniques to include in HEPData the features obtained from the classifiers. Finally, to gain a semi-quantitative estimate of phase space/parameter gaps, simpler analyses that are easily recastable are useful. 

For the early career session, particularly for the SVJ-l, it was noted that the use of single-lepton triggers is limited by the isolation requirements, and \pt thresholds, mainly from the Level1 side. PEB is preferred rather than lower di-muon triggers, because of high rates of fake muons specifically for ATLAS. An alternate trigger approach would be three lepton triggers, if the trigger bandwidth permits.

There were some follow up discussions on the definition of \rinv for SVJ-l, that is summarised in the list below:
\begin{itemize}
    \item If \rinv is forced as an input parameter, it can be taken at face value, and is easier to understand for theorists
    \item If \rinv is made an output parameter, then accurate branching ratios should be taken into account. However, the downside to this approach is that the \rinv becomes much more model-dependent and difficult to reinterpret.
    \item If \rinv can potentially be used as a measure of how to characterise angular distribution of jets in the $t$-channel with respect to \met, or shape estimation
    \item Invisible fraction of total energy for a given model is preferred, especially when multiple flavours of dark quarks is involved
\end{itemize}

Finally, there was a discussion on generation of the SVJ-l signatures. This has been studied for the first time in \cite{Cazzaniga:2022hxl,Beauchesne:2022phk}, however an other possibilities mentioned in the last talk of the session can be further explored. Additionally, it was mentioned that if \Nfdark is set to 3 in the hidden valley module, then $K_d \rightarrow \pi_d \pi_d$ due to weak interactions is triggered, and if there are flavour changing neutral currents, then $K_d \rightarrow \pi_d \gamma_d$ are also a possible generation mode for the signature. This leads to less leptons in final state, but potentially interesting physics.


\section{Acknowledgements}

 C.~Cazzaniga is supported by the Swiss National Science Fundation (SNFS) under the SNSF Eccellenza program. 
A. Garcia-Bellido's research is supported by the US Department of Energy award DE-SC0008475.
D.~Kar is supported by South Africa CERN research consortium.
S.~Kulkarni is supported by Austrian Science Fund research group funding FG1. P.~Schwaller is supported by the Cluster of Excellence “Precision Physics, Fundamental Interactions, and Structure of Matter” (PRISMA+ EXC 2118/1) funded by the German Research Foundation (DFG) within the German Excellence Strategy (Project No. 39083149).
Research by S.~Sinha is part of a project that has received funding from the European Research Council under the European Union’s Horizon 2020 research and innovation program (grant agreement 101002463). 
D.~Wilson-Edwards' research is supported by European Research Council grant REALDARK (grant agreement no. 101002463) and the Science and Technology Facilities Council, part of the UK Research and Innovation. 
J.~Zurita is supported by the Generalitat Valenciana (Spain) through the plan GenT program
(CIDEGENT/2019/068), by the Spanish Government (Agencia Estatal de Investigación)
and ERDF funds from European Commission (MCIN/AEI/10.13039/501100011033, Grant
No. PID2020-114473GB-I00). 

\appendix 
\section{Detailed summary of experimental talks}

\subsection{Search for s-channel semi-visible jets with CMS in Run 2 }
\textbf{Speaker: Aran Garcia-Bellido} \\ \\
The CMS experiment published in 2022 the first search for resonant production of ``semi-visible jets'' (SVJ) using 138~$\ifb$ of data from Run 2 \cite{CMS:2021dzg}. The theoretical model used for the search is based on the proposal in \cite{Cohen:2015toa,Cohen:2017pzm}, where 
a heavy leptophobic $Z^{'}$ boson from broken U(1) symmetry couples the Standard Model quarks (with coupling strength $g_q$) with dark quarks (with coupling strength $g_{\chi}$) belonging to a QCD-like Hidden sector. As mentioned in Section \ref{section:semi-visible jets pheno}, the dark quarks shower and hadronise in the hidden sector leading to dark bound states with degenerate masses $M_d$ (pseudo-scalar mesons $\pi_d$, and vector mesons $\rho_d$) that can be stable or can decay promptly back to Standard Model quarks. Unstable $\rho_d$ can decay to a pair of SM quarks of any flavor with equal probability,
while unstable $\pi_d$ must decay through a mass insertion, thus decays to the heaviest SM quarks kinematically accessible are preferred (usually b-quarks). The final signature is mainly characterized by at least two large jets with heavy-flavour content and $\cancel{E}_{\text{T}}$ aligned to one of them. The SVJ signal is simulated at leading order with the Hidden Valley model implemented with PYTHIA 8.226 for 2016 and PYTHIA 8.230 for 2017 and 2018~\cite{Sjostrand:2014zea}. Both the number of dark colours the number of dark flavours have been set to 2. The main parameters of the signal model have been listed in Section \ref{section:semi-visible jets pheno}, where the mediator mass in this case is the mass of the $Z^{'}$ boson $M_{Z'}$. For the reference parameters different values have been scanned: 19 values for  $M_{Z'} \in [1.5, 5.1]$~TeV, 12 values for $r_{\text{inv}} \in [0,1]$, 12 values for $M_d \in [1,100]$~GeV and 3 values for $\alpha_d$ (changing resulting multiplicities of dark hadrons from the shower and hadronization). Namely, $\alpha_{peak}$ is denoted as the value of the dark gauge coupling constant at 1 TeV for which the multiplicity of dark hadrons is maximised, while $\alpha_{low}$ and $\alpha_{high}$ are respectively $\frac{1}{2}\alpha_{peak}$ and $\frac{3}{2}\alpha_{peak}$. Three 2D scans changing $M_{Z'}$ has a function of the other parameters have been performed with a total 575 signal points.  
The $Z^{'}$ boson couplings have been chosen such that $g_q = 0.25$ and $g_{\chi} = 0.5$. With these choices, the production cross section, branching fraction,
and mediator width are compatible with the benchmark model recommended by the LHC DM Working Group~\cite{Boveia:2016mrp}. \\ \\
The search strategy consists in looking for the heavy mediator $Z^{'}$ producing a bump in the transverse mass spectrum of the di-jet system $M_T$. Two  main approaches have been followed: 1. inclusive search using only event-level variables which should give “model-independent” results, 2. BDT-based search: train a SVJ tagger using our
signal model, and explore the full sensitivity leveraging the expected differences in jet substructure between semi-visible jets and Standard model ones. \\ \\
The data is recorded with jet $p_{\text{T}}$ and HT triggers. At least two anti-$k_{\text{T}}$ with radius $ R = 0.8$, $p_{\text{T}} > 200$~GeV and $|\eta|<2.4$ are required. A selection on the transverse ratio is required $R_T =  \cancel{E}_{\text{T}}/M_{T}$ and used in the selection instead of $\cancel{E}_{\text{T}}$
in order to avoid the transverse mass sculpting and to identify events with invisible particles. The remaining t-channel QCD events are rejected requiring the pseudo-rapidity separation between the two highest $p_{\text{T}}$ jets to be $\Delta \eta (J1, J2) < 1.5$. Events with $M_T > 1.5$~TeV are selected to be in the fully efficient region for the triggers. In order to reduce the $t\bar{t}$ and $W(\ell \nu)+\text{jets}$ backgrounds, events containing
mini-isolated electrons or muons are vetoed. Events with anomalously high $\cancel{E}_{\text{T}}$ values can occur due to a variety of reconstruction failures, detector malfunctions, or other non-collision backgrounds. These events are rejected by custom filters tested in a dedicated control region at high $\Delta \eta (J1, J2)$. The final selection is completed requiring the minimum azimuthal separation between the two highest $p_{\text{T}}$ jets and the missing momentum $\cancel{E}_{\text{T}}$ to be $\Delta \phi_{\text{min}} <  0.8$. This requirement allows to further reject the electroweak backgrounds from $W/Z+\text{jets}$ and select a region of the phase space complementary to WIMPs searches. \\
On top of the previous selections, in the BDT-based search a jet-tagger is trained and applied to try to discriminate between SVJ and SM background jets using jet substructure variables. This tagger mainly employs 3 types of jet substructure variables related to: heavy object tagging (soft dropped mass, N-subjettinneses, energy correlation fucntions), quark-gluon discrimination (axis, generalized angularities) and flavour-based (energy fractions). The BDT has been trained with equal mix of QCD and $t\bar{t}$, and with a mixture of many signal models. To avoid mass sculpting, the background jets $p_{\text{T}}$ spectrum has been reweighted to match the signal. The tagger achieves overall very good performance with AUC $\sim 0.93-0.95$ with respect to all backgrounds.  \\ \\
The background estimation for both the inclusive and BDT-based strategy is performed via an analytic fit to $M_T$ data distribution. For the inclusive search strategy, a further categorization in $R_T$ is employed: the data are divided in a low-$R_T$ region  ($0.15<R_T<0.25$) and a high-$R_T$ region ($R_T>0.25$). Including the low-$R_T$ region improves the expected limit
by $\sim 60 \%$ and helps covering $r_{\text{inv}} \sim 0$ scenario. When the BDT is employed, subsets of the
low-$R_T$ and high-$R_T$ inclusive signal regions are selected by requiring that both jets in each event are
tagged as semi-visible. Events in which only one jet is tagged as semi-visible are not found to provide significant additional sensitivity. The results from the inclusive signal regions exclude observed (expected) values of up to $1.5 < M_{Z'} < 4.0$ TeV ($1.5 < M_{Z'} < 4.3$ TeV), with the widest exclusion range for models with $\alpha_{dark} = \alpha_{low}$. Depending on the $Z^{'}$ mass, $0.07 < r_{\text{inv}} < 0.53$ ($0.06 < r_{\text{inv}} < 0.57$) and all $M_d$ and $\alpha_{dark}$ variations considered are also observed (expected) to be excluded. The results from the BDT-based signal regions increase the observed (expected) excluded mediator mass range to $1.5 < M_{Z'} < 5.1$ TeV ($1.5 < M_{Z'} < 5.1$ TeV) for wide ranges of the other signal parameters. The range of observed (expected) excluded rinv values also increases to $0.01 < r_{\text{inv}} < 0.77$ ($0.01 < r_{\text{inv}} < 0.78$), and the $M_d$ and $\alpha_{dark}$ variations are excluded for a wider range of $Z^{'}$ masses.
\\ \\
Within the CMS experiment, new analysis on SVJ are ongoing trying to cover new event topologies, such as in the case of the t-channel production, as well as trying to access lower masses for the $Z^{'}$ in the boosted topology and using Data scouting. Moreover, anomaly detection techniques relying on autoencoders are under investigation for an unsupervised jet-tagger for semi-visible jets allowing to have a more model-independent search \cite{Canelli:2021aps}. Finally, new signatures for leptons-enriched semi-visible jets  have been proposed in \cite{Cazzaniga:2022hxl,Beauchesne:2022phk}, and a further analysis on these two signatures is in preparation.

\subsection{Search for emerging jets with CMS in Run 2}
\textbf{Speaker: Jannicke Pearkes} \\ \\
\label{section:emerging jets cms}
The CMS experiment published in 2019 the first search for ``emerging jets'' (EJ) using 16.1~$\ifb$ of data from 2016 \cite{CMS:2018bvr}. The theoretical model is based on the proposal in \cite{Schwaller:2015gea}, which includes a complex scalar mediator $X_{\text{DK}}$, an SU(3) color triplet in SM QCD, that can be pair produced via gluon fusion or quark-antiquark annihilation. Each mediator then decays to a dark quark $Q_\text{DK}$ and a SM quark. The dark quarks form dark pions that typically will have long lifetimes before they decay back to SM particles. So the final signature is four jets: two SM jets and two emerging jets (EJ). The EJ signal is simulated at leading order with the Hidden Valley model implemented with modified PYTHIA 8.212~\cite{Sjostrand:2014zea}. The number of dark colours is three, the number of dark flavours is set to 7, $\Lambda = m_{Q,\text{DK}}$, and $\Gamma_{X,DK}$ is 10~$\GeV$. All the $Q_\text{DK}$ are mass degenerate, and the meson masses are set such that: $m_{\pi,\text{DK}} = 0.5m_{Q,\text{DK}}$ and $m_{\rho,\text{DK}} = 2m_{Q,\text{DK}}$. These assumptions reduce the number of free parameters to three: the mediator mass $m_{X,DK}$, and the mass and decay length of the dark pion: $m_{\pi,\text{DK}}$ and $c\tau_{\pi,\text{DK}}$. 

The data is recorded with $\HT > 900~\GeV$ triggers and four anti-$k_{\text{T}}$ jets (R = 0.4) are required in the reconstruction with at least one track in each jet. The analysis uses four kinematic variables based on the track impact parameters to identify EJs with varying significance into six ``groups'', and then seven optimized selection ``sets'' to define signal and background enriched regions based on other event variables like $\HT$, the $\pt$ of the four jets, $\met$, the number of identified EJs corresponding to a given group. 

For a given $m_{\pi,\text{DK}}$, the analysis is most sensitive (has highest acceptance) for intermediate decay lengths ($25 < c\tau_{\pi,\text{DK}} < 100$~mm) and high mediator masses ($m_{X,DK} > 1200~\GeV$). For lower mediator masses, the trigger requirement lowers the acceptance; and for small decay lengths the dark pions decay very fast making them indistinguishable from SM QCD jets, while for large decay lengths the displaced tracks fall outside the tracking volume. Limits are set at 95\% confidence level excluding dark pion decay lengths between 5 and 225~mm for dark mediators with masses between 400 and 1250~$\GeV$. Decay lengths smaller than 5 and greater than 225~mm are also excluded in the lower part of this mass range. A paper is in preparation with the full Run-2 data which will have almost 10 times more data.

The CMS EJ group is currently exploring more dark QCD models in this signature, is implementing more targeted searches using machine learning to tag the jets, and is working to combine these results with the semivisible jets signature~\cite{CMS:2021dzg}. For Run 3 currently taking place, new triggers have become accessible that include displaced jets and anomaly detection triggers.

\subsection{Search for t-channel semi-visible jets with ATLAS in Run 2}
\label{section:semi-visible jets pheno}
\textbf{Speaker: Deepak Kar}

The search for t-channel Semi-visible jets \cite{ATLAS:2023swa} was performed with the full Run 2 dataset. For the theory model, the The Pythia8 HV dark coupling was set to be running at one-loop, the number of dark flavours ($N_{\mathrm{flav}}$) was set to 1, and the dark confinement scale ($\Lambda_D$) was set to $6.5$ TeV. It was raised in discussion during the workshop, that the choice of $N_{\mathrm{flav}} = 1$ could be theoretically problematic.

At leading order, two SVJs are back-to-back with the direction of the $\cancel{E}_{\text{T}}$ aligned with one of the jets - which is a signature dominated by dijet background processes. The addition of extra jets results in a boost in the cross section. A boost by additional jets leads to signatures with the $\cancel{E}_{\text{T}}$ not pointing necessarily in the direction of one of the two SVJs. Conversely, for multijet processes, the $\cancel{E}_{\text{T}}$ is typically aligned with one of the jets, as it usually arises due to mis-measured jets. Thus, these events are typically discarded in searches including jets and $\cancel{E}_{\text{T}}$. \\
\newline
Events in this analysis were selected based on the presence of two central jets, $\cancel{E}_{\text{T}}$ trigger, leading jet $p_{\text{T}} > 250$ GeV,  $H_{\text{T}} > 600$ GeV, $\cancel{E}_{\text{T}} > 600$ GeV, and jet closest to MET with $\Delta \phi < 2$. Three dedicated control regions were constructed for background estimation. Challenges arose from difficulties modelling signals with lower $\cancel{E}_{\text{T}}$, since they are buried by QCD signatures with fake $\cancel{E}_{\text{T}}$. To further address this, MET smearing techniques can be explored at the particle level (\textsc{HepData} \cite{Maguire_2017}, as well as \textsc{Rivet} \cite{Bierlich_2020} and \textsc{MadAnalysis} 5 \cite{Conte_2013} routines will soon be made available). \\
\newline 
While it is necessary to include benchmark models, it was also noted that if the model demonstrates sufficient versatility, a signature-driven search approach may be warranted. Direct connections do not always exist between our motivation and the specific signature, exemplified by cases such as R-Parity Violating SUSY, where MET is notably absent. As a community, we should decide whether we opt for reasonably sensible models and more signature driven searches, as opposed to a fully bottom-up theoretically driven choice of models. Namely, two options were outlined:

\begin{itemize}
    \item \textbf{Option 1:} Encourage the exploration of parameter choices like $N_c = 3$ and $N_{\mathrm{flav}} = 1$, and manipulating $r_{\text{inv}}$ to generate a wide range of values - with the caveat that we must explicitly define which observables and techniques should be avoided.
    \item \textbf{Option 2:} Contemplate experimental strategies first and then design models tailored to those strategies. For example, if all quarks are degenerate, and dark pions decay to heavy flavor (HF), the triggering mechanism becomes a significant part of the question. Alternatively, if cascade decays lead to b-quarks decaying into leptons, this demands a different approach to triggering. Therefore, identifying the desired signatures, and then creating models in alignment with these strategies, represents a distinct school of thought.
\end{itemize}

\subsection{Search for dark jet resonances with ATLAS in Run 2}
\label{section:djr_atlas}
\textbf{Speaker: Dilia Maria Portillo Quintero}

The search for dark jet resonances \cite{ATLAS-CONF-2023-047} was performed with the full Run 2 dataset recorded by ATLAS. Four benchmark models, introduced in \cite{Park_2019}, were considered in the search. Large radius jets were considered as they better encapsulated the double hadronisation procedure and resonance structure. To reduce the background and increase the analysis' sensitivity to the signal, dark jets were tagged using jet substructure information - namely the number of ungroomed tracks associated to a jet $n_{\mathrm{track}}$. A resonance was then searched for over the smoothly falling dijet invariant mass ($m_{jj}$) distribution. Challenges encountered during the analysis included defining a new observable to decorrelate $n_{\mathrm{track}}$ from $m_{jj}$, decision of whether $n_{\mathrm{track}}$ should be defined in data or MC, and estimating the background. 

In discussion, it was noted that as a first iteration, this search had limitations, for example in terms of the HV parameter choices and the discriminating variable - $n_{\mathrm{track}}$. The variable $n_{\mathrm{track}}$ is IRC unsafe and may not be suitable for the next iterations, suggestions for other variables with discriminating power are welcomed. Further, it was mentioned that if the dark hadrons had mass greater than $15 / 20$ GeV, new features can potentially show up in the tails of the $m_{jj}$ distributions. Probing subjets within large radius jets (a mostly model independent fact) could give insight into the nature of the jet, since these subjets are metastable hidden hadrons \cite{strassler2008phenomenology}. One could check the correlations between the hardest subjets to verify if it’s an actual signal, and as a community, investigating the masses of the subjets should remain in the agenda. This approach can be considered for discriminating between from jets orginating from quarks / gluons and unconventional jets.

\bibliographystyle{jhep}
\bibliography{references.bib}

\providecommand{\href}[2]{#2}\begingroup\raggedright\begin{thebibliography}{10}

\bibitem{Strassler:2006im}
M.~J. Strassler and K.~M. Zurek, \emph{{Echoes of a hidden valley at hadron
  colliders}},
  \href{http://dx.doi.org/10.1016/j.physletb.2007.06.055}{\emph{Phys. Lett. B}
  {\bfseries 651} (2007) 374--379},
  [\href{https://arxiv.org/abs/hep-ph/0604261}{{\ttfamily hep-ph/0604261}}].

\bibitem{Han:2007ae}
T.~Han, Z.~Si, K.~M. Zurek and M.~J. Strassler, \emph{{Phenomenology of hidden
  valleys at hadron colliders}},
  \href{http://dx.doi.org/10.1088/1126-6708/2008/07/008}{\emph{JHEP} {\bfseries
  07} (2008) 008}, [\href{https://arxiv.org/abs/0712.2041}{{\ttfamily
  0712.2041}}].

\bibitem{Juknevich:2009ji}
J.~E. Juknevich, D.~Melnikov and M.~J. Strassler, \emph{{A Pure-Glue Hidden
  Valley I. States and Decays}},
  \href{http://dx.doi.org/10.1088/1126-6708/2009/07/055}{\emph{JHEP} {\bfseries
  07} (2009) 055}, [\href{https://arxiv.org/abs/0903.0883}{{\ttfamily
  0903.0883}}].

\bibitem{ATLAS:2023jyp}
{ATLAS Collaboration}, \emph{{Search for dark photons in rare $Z$ boson decays
  with the ATLAS detector}},
  \href{https://arxiv.org/abs/2306.07413}{{\ttfamily 2306.07413}}.

\bibitem{Albouy:2022cin}
G.~Albouy et~al., \emph{{Theory, phenomenology, and experimental avenues for
  dark showers: a Snowmass 2021 report}},
  \href{http://dx.doi.org/10.1140/epjc/s10052-022-11048-8}{\emph{Eur. Phys. J.
  C} {\bfseries 82} (2022) 1132},
  [\href{https://arxiv.org/abs/2203.09503}{{\ttfamily 2203.09503}}].

\bibitem{Schwaller:2015gea}
P.~Schwaller, D.~Stolarski and A.~Weiler, \emph{{Emerging Jets}},
  \href{http://dx.doi.org/10.1007/JHEP05(2015)059}{\emph{JHEP} {\bfseries 05}
  (2015) 059}, [\href{https://arxiv.org/abs/1502.05409}{{\ttfamily
  1502.05409}}].

\bibitem{Mies:2020mzw}
H.~Mies, C.~Scherb and P.~Schwaller, \emph{{Collider constraints on dark
  mediators}}, \href{http://dx.doi.org/10.1007/JHEP04(2021)049}{\emph{JHEP}
  {\bfseries 04} (2021) 049},
  [\href{https://arxiv.org/abs/2011.13990}{{\ttfamily 2011.13990}}].

\bibitem{Renner:2018fhh}
S.~Renner and P.~Schwaller, \emph{{A flavoured dark sector}},
  \href{http://dx.doi.org/10.1007/JHEP08(2018)052}{\emph{JHEP} {\bfseries 08}
  (2018) 052}, [\href{https://arxiv.org/abs/1803.08080}{{\ttfamily
  1803.08080}}].

\bibitem{Carmona:2021seb}
A.~Carmona, C.~Scherb and P.~Schwaller, \emph{{Charming ALPs}},
  \href{http://dx.doi.org/10.1007/JHEP08(2021)121}{\emph{JHEP} {\bfseries 08}
  (2021) 121}, [\href{https://arxiv.org/abs/2101.07803}{{\ttfamily
  2101.07803}}].

\bibitem{Carmona:2022jid}
A.~Carmona, F.~Elahi, C.~Scherb and P.~Schwaller, \emph{{The ALPs from the top:
  searching for long lived axion-like particles from exotic top decays}},
  \href{http://dx.doi.org/10.1007/JHEP07(2022)122}{\emph{JHEP} {\bfseries 07}
  (2022) 122}, [\href{https://arxiv.org/abs/2202.09371}{{\ttfamily
  2202.09371}}].

\bibitem{Cohen:2020afv}
T.~Cohen, J.~Doss and M.~Freytsis, \emph{{Jet Substructure from Dark Sector
  Showers}}, \href{http://dx.doi.org/10.1007/JHEP09(2020)118}{\emph{JHEP}
  {\bfseries 09} (2020) 118},
  [\href{https://arxiv.org/abs/2004.00631}{{\ttfamily 2004.00631}}].

\bibitem{Batz:2023zef}
A.~Batz, T.~Cohen, D.~Curtin, C.~Gemmell and G.~D. Kribs, \emph{{Dark Sector
  Glueballs at the LHC}},  \href{https://arxiv.org/abs/2310.13731}{{\ttfamily
  2310.13731}}.

\bibitem{Dreyer:2018nbf}
F.~A. Dreyer, G.~P. Salam and G.~Soyez, \emph{{The Lund Jet Plane}},
  \href{http://dx.doi.org/10.1007/JHEP12(2018)064}{\emph{JHEP} {\bfseries 12}
  (2018) 064}, [\href{https://arxiv.org/abs/1807.04758}{{\ttfamily
  1807.04758}}].

\bibitem{Cohen:2023mya}
T.~Cohen, J.~Roloff and C.~Scherb, \emph{{Dark sector showers in the Lund jet
  plane}}, \href{http://dx.doi.org/10.1103/PhysRevD.108.L031501}{\emph{Phys.
  Rev. D} {\bfseries 108} (2023) L031501},
  [\href{https://arxiv.org/abs/2301.07732}{{\ttfamily 2301.07732}}].

\bibitem{CMS:2018bvr}
{CMS Collaboration}, \emph{{Search for new particles decaying to a jet and an
  emerging jet}}, \href{http://dx.doi.org/10.1007/JHEP02(2019)179}{\emph{JHEP}
  {\bfseries 02} (2019) 179},
  [\href{https://arxiv.org/abs/1810.10069}{{\ttfamily 1810.10069}}].

\bibitem{CMS:2021dzg}
{\scshape CMS} collaboration, A.~Tumasyan et~al., \emph{{Search for resonant
  production of strongly coupled dark matter in proton-proton collisions at 13
  TeV}}, \href{http://dx.doi.org/10.1007/JHEP06(2022)156}{\emph{JHEP}
  {\bfseries 06} (2022) 156},
  [\href{https://arxiv.org/abs/2112.11125}{{\ttfamily 2112.11125}}].

\bibitem{ATLAS:2023swa}
{ATLAS Collaboration}, \emph{{Search for non-resonant production of
  semi-visible jets using Run\textasciitilde{}2 data in ATLAS}},
  \href{https://arxiv.org/abs/2305.18037}{{\ttfamily 2305.18037}}.

\bibitem{ATLAS-CONF-2023-047}
{ATLAS Collaboration}, ``{Search for resonant production of dark quarks in the
  di-jet final state with the ATLAS detector}.'' {ATLAS-CONF-2023-047}, 2023.

\bibitem{Strassler:2008fv}
M.~J. Strassler, \emph{{On the Phenomenology of Hidden Valleys with Heavy
  Flavor}},  \href{https://arxiv.org/abs/0806.2385}{{\ttfamily 0806.2385}}.

\bibitem{Maguire:2017ypu}
E.~Maguire, L.~Heinrich and G.~Watt, \emph{{HEPData: a repository for high
  energy physics data}},
  \href{http://dx.doi.org/10.1088/1742-6596/898/10/102006}{\emph{J. Phys. Conf.
  Ser.} {\bfseries 898} (2017) 102006},
  [\href{https://arxiv.org/abs/1704.05473}{{\ttfamily 1704.05473}}].

\bibitem{Sjostrand:2014zea}
T.~Sj\"ostrand, S.~Ask, J.~R. Christiansen, R.~Corke, N.~Desai, P.~Ilten
  et~al., \emph{{An introduction to PYTHIA 8.2}},
  \href{http://dx.doi.org/10.1016/j.cpc.2015.01.024}{\emph{Comput. Phys.
  Commun.} {\bfseries 191} (2015) 159--177},
  [\href{https://arxiv.org/abs/1410.3012}{{\ttfamily 1410.3012}}].

\bibitem{Spergel:1999mh}
D.~N. Spergel and P.~J. Steinhardt, \emph{{Observational evidence for
  selfinteracting cold dark matter}},
  \href{http://dx.doi.org/10.1103/PhysRevLett.84.3760}{\emph{Phys. Rev. Lett.}
  {\bfseries 84} (2000) 3760--3763},
  [\href{https://arxiv.org/abs/astro-ph/9909386}{{\ttfamily
  astro-ph/9909386}}].

\bibitem{Oh_2011}
S.-H. Oh, C.~Brook, F.~Governato, E.~Brinks, L.~Mayer, W.~J.~G. de~Blok et~al.,
  \emph{{THE} {CENTRAL} {SLOPE} {OF} {DARK} {MATTER} {CORES} {IN} {DWARF}
  {GALAXIES}: {SIMULATIONS} {VERSUS} {THINGS}},
  \href{http://dx.doi.org/10.1088/0004-6256/142/1/24}{\emph{The Astronomical
  Journal} {\bfseries 142} (jun, 2011) 24}.

\bibitem{Rocha:2012jg}
M.~Rocha, A.~H.~G. Peter, J.~S. Bullock, M.~Kaplinghat, S.~Garrison-Kimmel,
  J.~Onorbe et~al., \emph{{Cosmological Simulations with Self-Interacting Dark
  Matter I: Constant Density Cores and Substructure}},
  \href{http://dx.doi.org/10.1093/mnras/sts514}{\emph{Mon. Not. Roy. Astron.
  Soc.} {\bfseries 430} (2013) 81--104},
  [\href{https://arxiv.org/abs/1208.3025}{{\ttfamily 1208.3025}}].

\bibitem{Hochberg:2014dra}
Y.~Hochberg, E.~Kuflik, T.~Volansky and J.~G. Wacker, \emph{{Mechanism for
  Thermal Relic Dark Matter of Strongly Interacting Massive Particles}},
  \href{http://dx.doi.org/10.1103/PhysRevLett.113.171301}{\emph{Phys. Rev.
  Lett.} {\bfseries 113} (2014) 171301},
  [\href{https://arxiv.org/abs/1402.5143}{{\ttfamily 1402.5143}}].

\bibitem{Bernreuther:2019pfb}
E.~Bernreuther, F.~Kahlhoefer, M.~Kr\"amer and P.~Tunney, \emph{{Strongly
  interacting dark sectors in the early Universe and at the LHC through a
  simplified portal}},
  \href{http://dx.doi.org/10.1007/JHEP01(2020)162}{\emph{JHEP} {\bfseries 01}
  (2020) 162}, [\href{https://arxiv.org/abs/1907.04346}{{\ttfamily
  1907.04346}}].

\bibitem{Belanger:2004yn}
G.~Belanger, F.~Boudjema, A.~Pukhov and A.~Semenov, \emph{{micrOMEGAs: Version
  1.3}}, \href{http://dx.doi.org/10.1016/j.cpc.2005.12.005}{\emph{Comput. Phys.
  Commun.} {\bfseries 174} (2006) 577--604},
  [\href{https://arxiv.org/abs/hep-ph/0405253}{{\ttfamily hep-ph/0405253}}].

\bibitem{Buckley:2021neu}
A.~Buckley et~al., \emph{{Testing new physics models with global comparisons to
  collider measurements: the Contur toolkit}},
  \href{http://dx.doi.org/10.21468/SciPostPhysCore.4.2.013}{\emph{SciPost Phys.
  Core} {\bfseries 4} (2021) 013},
  [\href{https://arxiv.org/abs/2102.04377}{{\ttfamily 2102.04377}}].

\bibitem{Butterworth:2016sqg}
J.~M. Butterworth, D.~Grellscheid, M.~Kr\"amer, B.~Sarrazin and D.~Yallup,
  \emph{{Constraining new physics with collider measurements of Standard Model
  signatures}}, \href{http://dx.doi.org/10.1007/JHEP03(2017)078}{\emph{JHEP}
  {\bfseries 03} (2017) 078},
  [\href{https://arxiv.org/abs/1606.05296}{{\ttfamily 1606.05296}}].

\bibitem{Bierlich:2019rhm}
C.~Bierlich et~al., \emph{{Robust Independent Validation of Experiment and
  Theory: Rivet version 3}},
  \href{http://dx.doi.org/10.21468/SciPostPhys.8.2.026}{\emph{SciPost Phys.}
  {\bfseries 8} (2020) 026},
  [\href{https://arxiv.org/abs/1912.05451}{{\ttfamily 1912.05451}}].

\bibitem{Kribs:2018ilo}
G.~D. Kribs, A.~Martin, B.~Ostdiek and T.~Tong, \emph{{Dark Mesons at the
  LHC}}, \href{http://dx.doi.org/10.1007/JHEP07(2019)133}{\emph{JHEP}
  {\bfseries 07} (2019) 133},
  [\href{https://arxiv.org/abs/1809.10184}{{\ttfamily 1809.10184}}].

\bibitem{Butterworth:2021jto}
J.~M. Butterworth, L.~Corpe, X.~Kong, S.~Kulkarni and M.~Thomas, \emph{{New
  sensitivity of LHC measurements to composite dark matter models}},
  \href{http://dx.doi.org/10.1103/PhysRevD.105.015008}{\emph{Phys. Rev. D}
  {\bfseries 105} (2022) 015008},
  [\href{https://arxiv.org/abs/2105.08494}{{\ttfamily 2105.08494}}].

\bibitem{ATLAS:2023vgr}
{ATLAS Collaboration}, ``{Search for dark mesons decaying to top and bottom
  quarks with the ATLAS detector in 140 fb$^{-1}$ of proton-proton collisions
  at $\sqrt{s}=13~$TeV}.'' {ATLAS-CONF-2023-021}, 2023.

\bibitem{GAMBIT:2023yih}
{\scshape GAMBIT} collaboration, V.~Ananyev et~al., \emph{{Collider constraints
  on electroweakinos in the presence of a light gravitino}},
  \href{http://dx.doi.org/10.1140/epjc/s10052-023-11574-z}{\emph{Eur. Phys. J.
  C} {\bfseries 83} (2023) 493},
  [\href{https://arxiv.org/abs/2303.09082}{{\ttfamily 2303.09082}}].

\bibitem{LHCReinterpretationForum:2020xtr}
{\scshape LHC Reinterpretation Forum} collaboration, W.~Abdallah et~al.,
  \emph{{Reinterpretation of LHC Results for New Physics: Status and
  Recommendations after Run 2}},
  \href{http://dx.doi.org/10.21468/SciPostPhys.9.2.022}{\emph{SciPost Phys.}
  {\bfseries 9} (2020) 022},
  [\href{https://arxiv.org/abs/2003.07868}{{\ttfamily 2003.07868}}].

\bibitem{Bailey:2022tdz}
S.~Bailey et~al., \emph{{Data and Analysis Preservation, Recasting, and
  Reinterpretation}},  \href{https://arxiv.org/abs/2203.10057}{{\ttfamily
  2203.10057}}.

\bibitem{Carrasco:2023loy}
J.~Carrasco and J.~Zurita, \emph{{Emerging jet probes of strongly interacting
  dark sectors}},  \href{https://arxiv.org/abs/2307.04847}{{\ttfamily
  2307.04847}}.

\bibitem{Kar:2022hxn}
D.~Kar and S.~Sinha, \emph{{2B or not 2B, a study of bottom-quark-philic
  semi-visible jets}},  \href{https://arxiv.org/abs/2207.01885}{{\ttfamily
  2207.01885}}.

\bibitem{Cazzaniga:2022hxl}
C.~Cazzaniga and A.~de~Cosa, \emph{{Leptons lurking in semi-visible jets at the
  LHC}}, \href{http://dx.doi.org/10.1140/epjc/s10052-022-10775-2}{\emph{Eur.
  Phys. J. C} {\bfseries 82} (2022) 793},
  [\href{https://arxiv.org/abs/2206.03909}{{\ttfamily 2206.03909}}].

\bibitem{Beauchesne:2022phk}
H.~Beauchesne, C.~Cazzaniga, A.~de~Cosa, C.~Doglioni, T.~Fitschen, G.~G.
  di~Cortona et~al., \emph{{Uncovering tau leptons-enriched semi-visible jets
  at the LHC}},
  \href{http://dx.doi.org/10.1140/epjc/s10052-023-11775-6}{\emph{Eur. Phys. J.
  C} {\bfseries 83} (2023) 599},
  [\href{https://arxiv.org/abs/2212.11523}{{\ttfamily 2212.11523}}].

\bibitem{ATLAS:2022yru}
{ATLAS Collaboration}, ``{SimpleAnalysis: Truth-level Analysis Framework}.''
  {ATL-PHYS-PUB-2022-017}, 2022.

\bibitem{Simko:2018zzz}
T.~\v{S}imko, L.~Heinrich, H.~Hirvonsalo, D.~Kousidis and D.~Rodr\'\i{}guez,
  \emph{{REANA: A System for Reusable Research Data Analyses}},
  \href{http://dx.doi.org/10.1051/epjconf/201921406034}{\emph{EPJ Web Conf.}
  {\bfseries 214} (2019) 06034}.

\bibitem{Buckley:2019stt}
A.~Buckley, D.~Kar and K.~Nordstr\"om, \emph{{Fast simulation of detector
  effects in Rivet}},
  \href{http://dx.doi.org/10.21468/SciPostPhys.8.2.025}{\emph{SciPost Phys.}
  {\bfseries 8} (2020) 025},
  [\href{https://arxiv.org/abs/1910.01637}{{\ttfamily 1910.01637}}].

\bibitem{Cohen:2015toa}
T.~Cohen, M.~Lisanti and H.~K. Lou, \emph{{Semivisible Jets: Dark Matter
  Undercover at the LHC}},
  \href{http://dx.doi.org/10.1103/PhysRevLett.115.171804}{\emph{Phys. Rev.
  Lett.} {\bfseries 115} (2015) 171804},
  [\href{https://arxiv.org/abs/1503.00009}{{\ttfamily 1503.00009}}].

\bibitem{Cohen:2017pzm}
T.~Cohen, M.~Lisanti, H.~K. Lou and S.~Mishra-Sharma, \emph{{LHC Searches for
  Dark Sector Showers}},
  \href{http://dx.doi.org/10.1007/JHEP11(2017)196}{\emph{JHEP} {\bfseries 11}
  (2017) 196}, [\href{https://arxiv.org/abs/1707.05326}{{\ttfamily
  1707.05326}}].

\bibitem{Boveia:2016mrp}
A.~Boveia et~al., \emph{{Recommendations on presenting LHC searches for missing
  transverse energy signals using simplified $s$-channel models of dark
  matter}}, \href{http://dx.doi.org/10.1016/j.dark.2019.100365}{\emph{Phys.
  Dark Univ.} {\bfseries 27} (2020) 100365},
  [\href{https://arxiv.org/abs/1603.04156}{{\ttfamily 1603.04156}}].

\bibitem{Canelli:2021aps}
F.~Canelli, A.~de~Cosa, L.~L. Pottier, J.~Niedziela, K.~Pedro and M.~Pierini,
  \emph{{Autoencoders for semivisible jet detection}},
  \href{http://dx.doi.org/10.1007/JHEP02(2022)074}{\emph{JHEP} {\bfseries 02}
  (2022) 074}, [\href{https://arxiv.org/abs/2112.02864}{{\ttfamily
  2112.02864}}].

\bibitem{Maguire_2017}
E.~Maguire, L.~Heinrich and G.~Watt, \emph{{HEPData}: a repository for high
  energy physics data},
  \href{http://dx.doi.org/10.1088/1742-6596/898/10/102006}{\emph{Journal of
  Physics: Conference Series} {\bfseries 898} (oct, 2017) 102006}.

\bibitem{Bierlich_2020}
C.~Bierlich, A.~Buckley, J.~Butterworth, C.~H. Christensen, L.~Corpe,
  D.~Grellscheid et~al., \emph{Robust independent validation of experiment and
  theory: Rivet version 3},
  \href{http://dx.doi.org/10.21468/scipostphys.8.2.026}{\emph{{SciPost}
  Physics} {\bfseries 8} (feb, 2020) }.

\bibitem{Conte_2013}
E.~Conte, B.~Fuks and G.~Serret, \emph{{MadAnalysis}~5, a user-friendly
  framework for collider phenomenology},
  \href{http://dx.doi.org/10.1016/j.cpc.2012.09.009}{\emph{Computer Physics
  Communications} {\bfseries 184} (jan, 2013) 222--256}.

\bibitem{Park_2019}
M.~Park and M.~Zhang, \emph{Tagging a jet from a dark sector with jet
  substructures at colliders},
  \href{http://dx.doi.org/10.1103/physrevd.100.115009}{\emph{Physical Review D}
  {\bfseries 100} (dec, 2019) }.

\bibitem{strassler2008phenomenology}
M.~J. Strassler, \emph{On the phenomenology of hidden valleys with heavy
  flavor},  2008.

\end{thebibliography}\endgroup
\end{document}